# New Families of LDPC Block Codes Formed by Terminating Irregular Protograph-Based LDPC Convolutional Codes

David G. M. Mitchell*, Michael Lentmaier†, and Daniel J. Costello, Jr.*
*Dept. of Electrical Engineering, University of Notre Dame, Indiana, USA,
{david.mitchell, costello.2}@nd.edu
†Vodafone Chair Mobile Communications Systems, Dresden University of Technology, Dresden, Germany,
michael.lentmaier@ifn.et.tu-dresden.de

*Abstract*— **In this paper, we present a method of constructing new families of LDPC block code ensembles formed by terminating irregular protograph-based LDPC convolutional codes. Using the accumulate-repeat-by-4-jagged-accumulate (AR4JA) protograph as an example, a density evolution analysis for the binary erasure channel shows that this flexible design technique gives rise to a large selection of LDPC block code ensembles with varying code rates and thresholds close to capacity. Further, by means of an asymptotic weight enumerator analysis, we show that all the ensembles in this family also have minimum distance that grows linearly with block length, i.e., they are asymptotically good.**

## I. Introduction

The notion of *degree distribution* is an important factor in the design of irregular low-density parity-check (LDPC) codes. The performance of a belief propagation (BP) decoder for LDPC codes is closely related to the variable and check node degrees in the Tanner graph of the code. In order to improve decoder performance, irregular code ensembles with a variety of node degrees have been proposed (see, e.g., [1]). For the binary erasure channel (BEC), capacity achieving sequences of codes with a vanishing gap between the threshold and the Shannon limit $\varepsilon_{sh} = 1 - R$ have been found in [2]. The node degree distribution also influences the minimum distance properties of the code ensemble. For example, $(J, K)$-regular LDPC code ensembles [3] with constant node degrees have minimum distance that grows linearly with block length for $J > 2$, i.e., they are *asymptotically good*; however, they also have comparitively poor iterative decoding thresholds.

LDPC codes based on a *protograph* [4] form a subclass of multi-edge type codes that have been shown in the literature to have many desirable features, such as good iterative decoding thresholds and, for suitably-designed protographs, linear minimum distance growth (see, e.g., [5]). So-called *asymptotically regular* LDPC block code ensembles [6] are formed by terminating $(J, K)$-regular protograph-based LDPC convolutional codes. This construction method results in LDPC block code ensembles with substantially better thresholds than those of $(J, K)$-regular LDPC block code ensembles, despite the fact that the ensembles are almost regular (see, e.g., [6]). As the termination length tends to infinity, it is further observed that the iterative decoding thresholds of these asymptotically good ensembles approach the optimal maximum a posteriori probability (MAP) decoding thresholds of the corresponding LDPC block code ensembles. Recently, this property has been proven analytically in [7] for the BEC considering some slightly modified ensembles. These codes were also shown to be asymptotically good in [8].

This work was partially supported by NSF Grant CCF08-30650 and NASA Grant NNX09AI66G.

In this paper, we extend the results of [8] to consider the effect of terminating irregular protograph-based LDPC convolutional code ensembles. As an example, we use an accumulate-repeat-by-4-jagged-accumulate (AR4JA) protograph [5] to construct an irregular LDPC convolutional code ensemble by means of an edge spreading technique over component matrices. By design, this LDPC convolutional code ensemble has the same degree distribution and rate as the AR4JA LDPC block code ensemble. Terminating this ensemble gives rise to a family of irregular LDPC block code ensembles with flexible code rates defined by a termination factor $L$. Using density evolution analysis for these ensembles over the BEC, we show that this flexible design technique gives rise to a large selection of code ensembles with varying code rates and thresholds close to capacity. Further, by means of an asymptotic weight enumerator analysis [5], we show that all the ensembles in this family are asymptotically good.

## II. Analysis of protograph-based LDPC codes

A protograph is a small bipartite graph $B = (V, C, E)$ that connects a set of $n_v$ variable nodes $V = \{v_0, \ldots, v_{n_v-1}\}$ to a set of $n_c$ check nodes $C = \{c_0, \ldots, c_{n_c-1}\}$ by a set of edges $E$. The protograph can be represented by a parity-check or *base* biadjacency matrix $\mathbf{B}$, where $\mathbf{B}_{j,k}$ is taken to be the number of edges connecting variable node $v_k$ to check node $c_j$. As an example, Figure 1 shows the accumulate-repeat-jagged-accumulate (ARJA) protograph [5] and its associated base matrix $\mathbf{B}$.

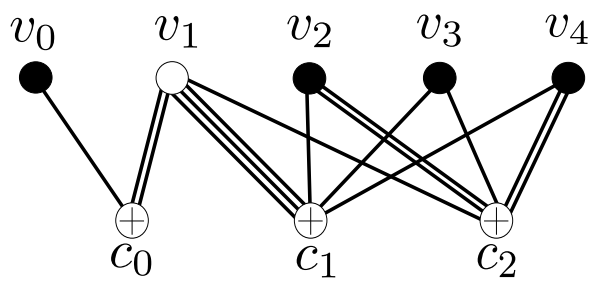

$$\mathbf{B} = \begin{bmatrix} 1 & 2 & 0 & 0 & 0 \\ 0 & 3 & 1 & 1 & 1 \\ 0 & 1 & 2 & 1 & 2 \end{bmatrix}$$

Fig. 1: The ARJA protograph and associated base matrix $\mathbf{B}$.

The ARJA protograph has multiple repeated edges between $V$ and $C$. In addition, as illustrated by the undarkened circle, variable node $v_1$ is punctured.

An ensemble of protograph-based LDPC block codes can be created from a base matrix $\mathbf{B}$ using the *copy-and-permute* operation [4]. A parity-check matrix $\mathbf{H}$ from the ensemble of protograph-based LDPC block codes can then be obtained by replacing ones with an $N \times N$ permutation matrix and zeros with the $N \times N$ all zero matrix in the base matrix $\mathbf{B}$. In the case when a variable node and a check node are connected by $r$ repeated edges, the associated entry in $\mathbf{B}$ equals $r$ and the corresponding block in $\mathbf{H}$ consists of a summation of $r$ $N \times N$ permutation matrices. The *ensemble* is defined as the set of all possible parity-check matrices $\mathbf{H}$ that can be formed using this method.

By construction, every code in the resulting ensemble has the same node degrees and structure. The ensemble rate is given as $R = (n_v - n_c)/u$, where $u$ is the number of variable nodes connected to the channel. In addition, the sparsity condition of an LDPC matrix is satisfied for large $N$. The code created by applying the copy-and-permute operation to an $n_c \times n_v$ protograph base matrix $\mathbf{B}$ has block length $n = Nn_v$.

### A. Density evolution for protograph-based ensembles

Since every member of the protograph-based ensemble preserves the structure of the base protograph, density evolution analysis for the resulting codes can be performed within the protograph. In this paper, we assume that belief propagation (BP) decoding is performed after transmission over a BEC with erasure probability $\varepsilon$. Let $p^{(i)}$ denote the probability that the incoming message in the previous update along an edge of an arbitrary check node is an erasure. Then the *density evolution threshold* of an ensemble is defined as the maximal value of the channel parameter $\varepsilon$ for which $p^{(i)}$ converges to zero for all edges as the number of iterations $i$ tends to infinity.

### B. Protograph weight enumeration

The preserved structure of members of a protograph-based LDPC code ensemble also facilitates the calculation of average weight enumerators. An *ensemble average weight enumerator* $A_d$ tells us that, given a particular Hamming weight $d$, a typical member of the ensemble has $A_d$ codewords with Hamming weight $d$. Combinatorial techniques for calculating enumerators for protograph-based ensembles have been presented in [5] and [9]. The weight enumerator $A_d$ can be analysed asymptotically to test if the ensemble is asymptotically good. If this is the case, then we can say that the majority of codes in the ensemble have minimum distance growing linearly at least as fast as $n\delta_{min}$, where $\delta_{min}$ is called the *minimum distance growth rate* of the code ensemble [5].

## III. Terminated LDPC convolutional codes

A rate $R = b/c$ binary LDPC convolutional code [10] can be defined as the set of infinite binary sequences $\mathbf{v}_{[-\infty,\infty]}$ that satisfy the equation $\mathbf{v}_{[-\infty,\infty]}\mathbf{H}^{\mathrm{T}}_{[-\infty,\infty]} = \mathbf{0}$, where

$$\mathbf{H}^T_{[-\infty,\infty]} = \begin{bmatrix} \ddots & & & \ddots & & \\ & \mathbf{H}_0^T(0) & \cdots & \mathbf{H}_{m_s}^T(m_s) & & \\ & & \ddots & & \ddots & \\ & & & \mathbf{H}_0^T(t) & \cdots & \mathbf{H}_{m_s}^T(t+m_s) \\ & & & & \ddots & & \ddots \end{bmatrix}$$

is the transposed parity-check matrix, also called the *syndrome former matrix*. The binary $(c-b) \times c$ submatrices $\mathbf{H}_i(t)$, $i = 0, 1, \cdots, m_s$, satisfy the conditions that $\mathbf{H}_{m_s}(t) \neq \mathbf{0}$ for at least one $t \in \mathbb{Z}$ and that $\mathbf{H}_0(t)$ has full rank for all $t$. We call $m_s$ the *syndrome former memory* and $\nu_s = (m_s+1) \cdot c$ the *decoding constraint length*. These parameters determine the width of the nonzero diagonal region of $\mathbf{H}_{[-\infty,\infty]}$. The sparsity of the parity-check matrix is ensured by demanding that its rows have Hamming weight much less than $\nu_s$. The code is said to be regular if its parity-check matrix $\mathbf{H}_{[-\infty,\infty]}$ has exactly $J$ ones in every column and $K$ ones in every row. The code is irregular if its row and column weights are not constant, and the degree distribution is used to characterize the variations of check and variable node degrees in the Tanner graph of the code.

### A. Constructing protograph-based LDPC convolutional codes

Analogously to block codes, an ensemble of LDPC convolutional codes can be constructed from a protograph. We proceed by forming an infinite base matrix composed of component $b_c \times b_v$ submatrices $\mathbf{B}_0, \mathbf{B}_1, \ldots, \mathbf{B}_{m_s}$ as follows:

$$\mathbf{B}_{[-\infty,\infty]} = \begin{bmatrix} \ddots & & \ddots & & \\ \mathbf{B}_{m_s} & \cdots & \mathbf{B}_0 & & \\ & \ddots & & \ddots & \\ & & \mathbf{B}_{m_s} & \cdots & \mathbf{B}_0 \\ & & & \ddots & & \ddots \end{bmatrix}. \quad (1)$$

The infinite Tanner graph associated with $\mathbf{B}_{[-\infty,\infty]}$ can be regarded as a *convolutional protograph*. An ensemble of time-varying LDPC convolutional codes can be formed from $\mathbf{B}_{[-\infty,\infty]}$ using the protograph construction method based on $N \times N$ permutation matrices described in Section II.

### B. Terminated LDPC convolutional codes

Suppose that we start the convolutional code with parity-check matrix defined in (1) at time $t = 0$ and terminate it after $L$ time instants. The resulting finite-length base matrix is given by

$$\mathbf{B}_{[0,L-1]} = \begin{bmatrix} \mathbf{B}_0 & & \\ \vdots & \ddots & \\ \mathbf{B}_{m_s} & & \mathbf{B}_0 \\ & \ddots & \vdots \\ & & \mathbf{B}_{m_s} \end{bmatrix}_{(L+m_s)b_c \times Lb_v}. \quad (2)$$

The matrix $\mathbf{B}_{[0,L-1]}$ can be considered as the base matrix of a terminated protograph-based LDPC convolutional code ensemble. Termination in this fashion results in a rate loss. Without puncturing, the design rate $R_L$ of the terminated code ensemble is equal to

$$R_L = 1 - \left(\frac{L+m_s}{L}\right)\frac{b_c}{b_v} = 1 - \left(\frac{L+m_s}{L}\right)(1-R),$$

where $R = 1 - Nb_c/Nb_v = 1 - b_c/b_v$ is the rate of the unterminated convolutional code ensemble. Note that, as the termination factor $L$ increases, the rate increases and approaches the rate of the unterminated convolutional code ensemble. The protograph-based LDPC block code ensemble associated with $\mathbf{B}_{[0,L-1]}$ can be studied using the analysis discussed in Section II. It has been shown in [6] and [8] that the BEC thresholds and minimum distance growth rates are highly dependent on the choice of component submatrices.

## IV. AR4JA-BASED TERMINATED LDPC CONVOLUTIONAL CODES

As an example of our design method, we construct families of protograph-based LDPC convolutional code ensembles based on the irregular AR4JA ensembles introduced in [5]. These convolutional code ensembles are then terminated following the procedure described in Section III-B, and we find the iterative decoding thresholds and minimum distance growth rates of the resulting LDPC block code ensembles.

### A. A family of terminated ARJA-based convolutional codes

The ARJA protograph [5] and associated base matrix are displayed in Figure 1. The ensemble defined by this protograph is of significant practical interest, since it has minimum distance growth rate $\delta_{min} = 0.0145$ and BEC iterative decoding threshold $\varepsilon^* = 0.4387$. Consider splitting $\mathbf{B}$ into component submatrices $\mathbf{B}_0$ and $\mathbf{B}_1$ of size $b_c \times b_v = 3 \times 5$ as follows:

$$\mathbf{B}_0 = \begin{bmatrix} 1 & 2 & 0 & 0 & 0 \\ 0 & 1 & 1 & 1 & 0 \\ 0 & 0 & 1 & 0 & 2 \end{bmatrix} \text{ and } \mathbf{B}_1 = \begin{bmatrix} 0 & 0 & 0 & 0 & 0 \\ 0 & 2 & 0 & 0 & 1 \\ 0 & 1 & 1 & 1 & 0 \end{bmatrix},$$

where we note that $\mathbf{B}_0+\mathbf{B}_1 = \mathbf{B}$. We can use these component submatrices to form a convolutional base matrix as in (1), where $m_s = 1$. The associated convolutional protograph is shown in Figure 2.

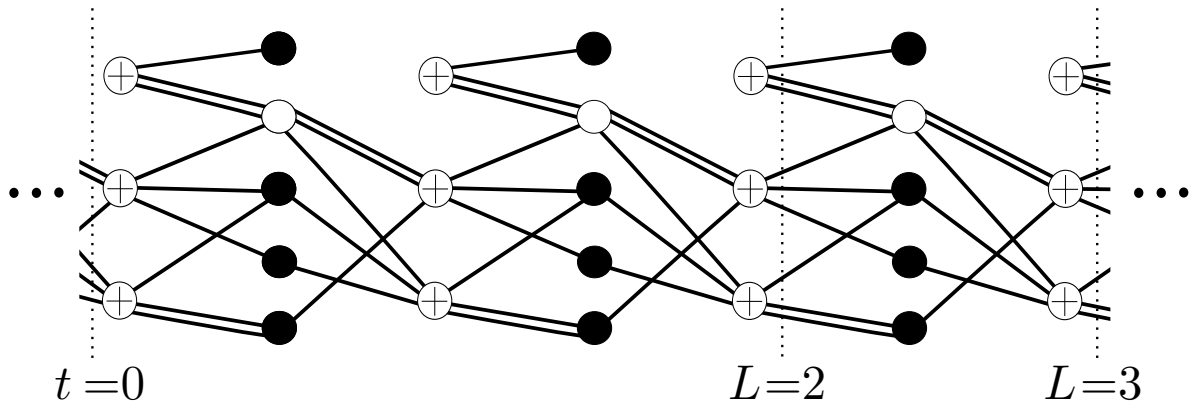

Fig. 2: The ARJA-based convolutional protograph defined in Section IV-A with termination markings for increasing $L$.

Note that the variable nodes associated with the second column of the component submatrices are punctured in accordance with the ARJA protograph. Using this construction method, the infinite convolutional protograph has the same degree distribution as the ARJA protograph and design rate $R = 1/2$. Spreading the edges of $\mathbf{B}$ over component submatrices in this fashion ensures that edges from variable nodes at time $t$ are spread among check nodes at times $t, t+1, \ldots, t+m_s$.

The convolutional protograph may then be terminated as shown in Figure 2, with the associated base matrix given by (2), for termination factors $L \geq 2$. In order to achieve linear distance growth, we have modified the construction presented in [6] to avoid degree one and two check nodes, so that the edges of the degree 3 check node $c_0$ are not split over the component submatrices. As a result of the all-zero row in $\mathbf{B}_1$, the terminated protograph associated with $\mathbf{B}_{[0,L-1]}$ has $n_c = (L+m_s)b_c - 1 = 3L+2$ check nodes and $n_v = Lb_v = 5L$ variable nodes. After puncturing, the design rate is

$$R_L = \frac{n_v - n_c}{u} = \frac{5L - (3L+2)}{4L} = \frac{L-1}{2L}.$$

Note that, while the terminated code ensembles approach the check node degree distribution of the ARJA ensemble as $L \to \infty$, for finite $L$ the terminated ensembles have a reduced fraction of degree 6 check nodes. For $L \geq 2$, the protograph has $L+4$ degree 3 check nodes and $2(L-1)$ degree 6 check nodes. By design, the variable node degree distribution remains constant for all $L$. The calculated minimum distance growth rates and BEC thresholds for these ensembles are given in Table I.

| $L$ | Rate $R_L$ | Growth rate $\delta_{min}^{(L)}$ | Scaled $u\delta_{min}^{(L)}$ | BEC threshold | Capacity $\varepsilon_{sh}$ | Gap to Capacity |
|---|---|---|---|---|---|---|
| 2 | 1/4 | 0.0946 | 0.757 | 0.6608 | 0.7500 | 0.0892 |
| 3 | 1/3 | 0.0461 | 0.553 | 0.5864 | 0.6667 | 0.0803 |
| 4 | 3/8 | 0.0306 | 0.490 | 0.5496 | 0.6250 | 0.0750 |
| 5 | 2/5 | 0.0234 | 0.469 | 0.5284 | 0.6000 | 0.0716 |
| 6 | 5/12 | 0.0192 | 0.462 | 0.5159 | 0.5833 | 0.0674 |
| 7 | 3/7 | 0.0164 | 0.461 | 0.5083 | 0.5714 | 0.0631 |
| 8 | 7/16 | 0.0144 | 0.461 | 0.5039 | 0.5625 | 0.0586 |
| 9 | 4/9 | 0.0128 | 0.461 | 0.5016 | 0.5556 | 0.0540 |
| 10 | 9/20 | 0.0115 | 0.461 | 0.5004 | 0.5500 | 0.0496 |
| $\infty$ | 1/2 | 0 | | 0.4996 | 0.5000 | 0.0004 |

TABLE I: Parameters for the terminated ARJA-based LDPC convolutional code ensembles.

As the termination factor $L \to \infty$, we observe that the minimum distance growth rate $\delta_{min}^{(L)} \to 0$.[1] This is consistent with similar results obtained for tail-biting LDPC convolutional code ensembles in [11]. We also observe from Table I that the scaled growth rates $u\delta_{min}^{(L)}$ converge to a fixed value as $L$ increases. A similar result was first observed in [12] for an ensemble of $(3,6)$-regular LDPC convolutional codes constructed from $N \times N$ permutation matrices, where it was shown that the scaled growth rates converged to a bound on the *free distance* growth rate of the unterminated LDPC convolutional code ensemble. This allows us to estimate the minimum distance growth rate $\delta_{min}^{(L)}$ for larger $L$ by dividing this value by the number of transmitted nodes in the protograph $u = 4L$.

In addition to the convergence of the scaled minimum distance growth rate with $L$, Table I also indicates that the BEC iterative decoding threshold converges to a constant value with $L$. As the termination factor $L$ increases, we also observe that the gap to capacity decreases. Since the distance growth rates decrease with $L$, this indicates the existence of a trade-off between distance growth rate and threshold. For this ensemble, the threshold approaches $\varepsilon^* = 0.4996$ as $L \to \infty$. This is very close to the Shannon limit $\varepsilon_{sh} = 0.5$ for rate $R_\infty = 1/2$. Importantly, the threshold does not further decay as the termination factor $L$ increases. This remarkable result was first observed empirically in [13] for $(J, 2J)$-regular ensembles constructed from $N \times N$ permutation matrices, and it was shown to be true for arbitrarily large $L$ in [14]. Recently, this phenomenon has also been observed for terminated protograph-based regular LDPC convolutional codes [6], [8].

[1] Using the techniques developed in [11], this convolutional code ensemble can be shown to be asymptotically good in the sense that the minimum *free distance* grows linearly with encoding constraint length.

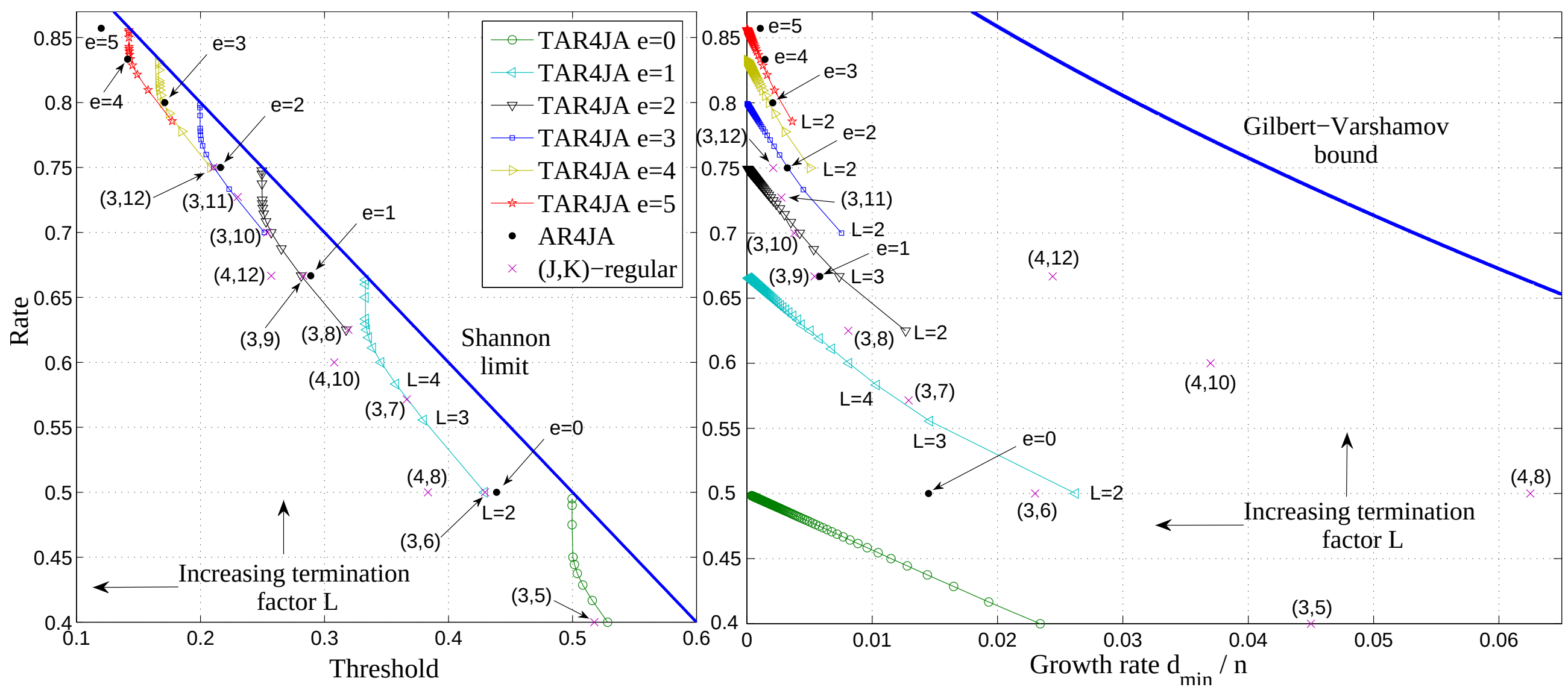

Fig. 5: Left: BEC iterative decoding threshold vs. code rate along with the Shannon limit. Right: minimum distance growth rates vs. code rate along with the Gilbert-Varshamov bound.

## B. *Terminated AR4JA-based LDPC convolutional codes*

The ARJA protograph can be extended to a family of AR4JA protographs [5] by adding $2e$ variable nodes of degree $4$ as shown in Fig. 3.

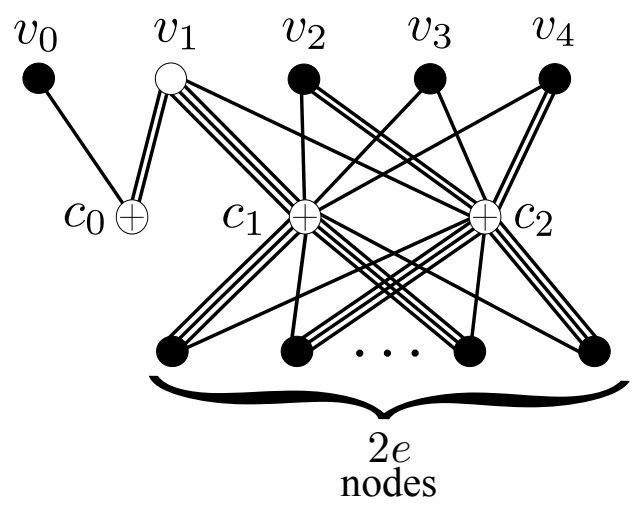

Fig. 3: The AR4JA protographs with extension parameter $e$.

This method of extension preserves linear minimum distance growth, and the LDPC code ensembles associated with this family of protographs have been shown to have good iterative decoding thresholds [5]. The design rate of the ensemble with extension parameter $e$ is $R = (1+e)/(2+e)$. Note that setting $e = 0$ results in the ARJA protograph.

Using the technique described above, we can construct an AR4JA-based LDPC convolutional code ensemble from (1) using component protographs $\mathbf{B}_0$ and $\mathbf{B}_1$ as shown in Fig. 4. Terminated AR4JA (TAR4JA) base matrices $\mathbf{B}_{[0,L-1]}$ can then be formed as in (2), where $\mathbf{B}_{[0,L-1]}$ is now of size $(3L+2) \times (5+2e)L$. Note that there are exactly $L$ punctured nodes in $\mathbf{B}_{[0,L-1]}$, where the nodes are punctured as indicated in Figure 4. As with the AR4JA LDPC block code, $e = 0$ corresponds to the terminated ARJA-based LDPC convolutional code ensemble described in Section IV-A. Then, for extension parameter $e$, the rate of any given terminated ensemble with termination factor $L \geq 2$ is given by

$$R_L = \frac{n_v - n_c}{m} = \frac{(5+2e)L - (3L+2)}{(4+2e)L} = \frac{(1+e)L-1}{(2+e)L}.$$

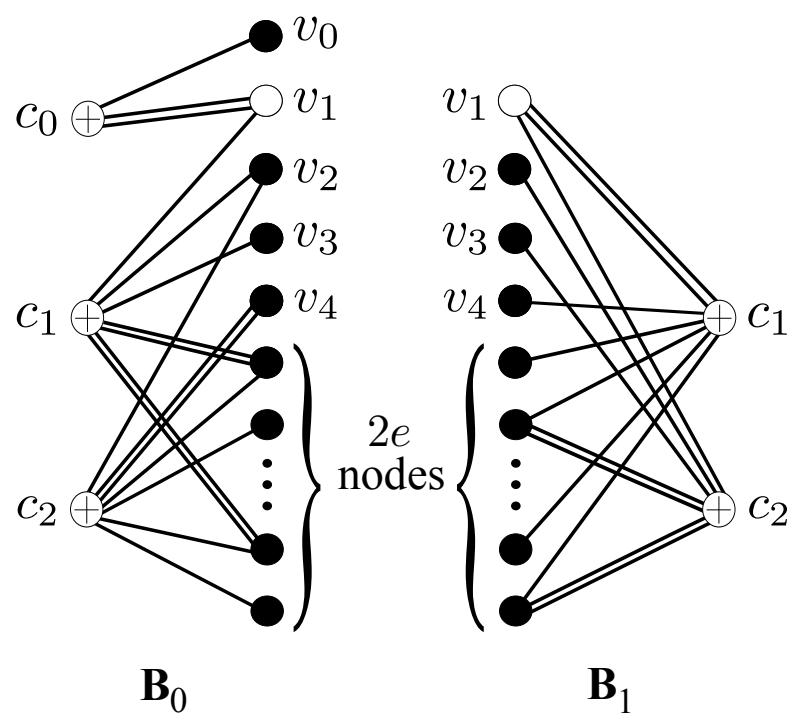

Fig. 4: The component protographs for the AR4JA-based LDPC convolutional code ensembles.

Figure 5 shows the results obtained for the TAR4JA ensembles, the AR4JA ensembles, and some $(J, K)$-regular ensembles. For $e = 1, \ldots, 5$, we again observe that the scaled minimum distance growth rates $u\delta_{min}^{(L)}$ of the TAR4JA ensembles converge as $L$ increases, which allows us to estimate the growth rates for $L \geq 10$.

For the TAR4JA ensembles with $e = 1, \ldots, 5$, we observe that, as with the $e = 0$ case, increasing the termination factor $L$ results in a family of codes with capacity approaching iterative decoding thresholds and declining minimum distance growth rates. For each family, the iterative decoding threshold converges to a value close to the Shannon limit for $R_\infty$ as $L$ gets large. The design rates $R_L$ of the TAR4JA ensembles overlap for increasing extension parameter $e$, allowing a large selection of asymptotically good codes to be obtained in the range $1/4 \leq R \leq 6/7$. The achieveable code rate can be increased by considering larger extension parameters $e$.

We also observe that the minimum distance growth rates of the TAR4JA ensembles for small termination factors $L$ typically exceed those of $(3, K)$-regular codes for $K \geq 6$. For the same extension parameter $e$ and large termination factors $L$, the TAR4JA ensembles have significantly better thresholds and less complexity then the AR4JA ensembles[2], but smaller distance growth rates and slightly lower code rates. Further, by increasing the extension parameter $e$, and for small termination factors $L$, the minimum distance growth rates of the TAR4JA ensembles are larger than those of the AR4JA ensemble with only a slightly worse threshold and some increase in complexity.

Figure 6 plots the minimum distance growth rates against the fractional gap to capacity $(\varepsilon_{sh} - \varepsilon^*)/\varepsilon_{sh}$ for the TAR4JA ensembles with termination factors $L = 2, \ldots, 10, 20, 50, 100$, the AR4JA ensembles, and some $(J, K)$-regular ensembles.

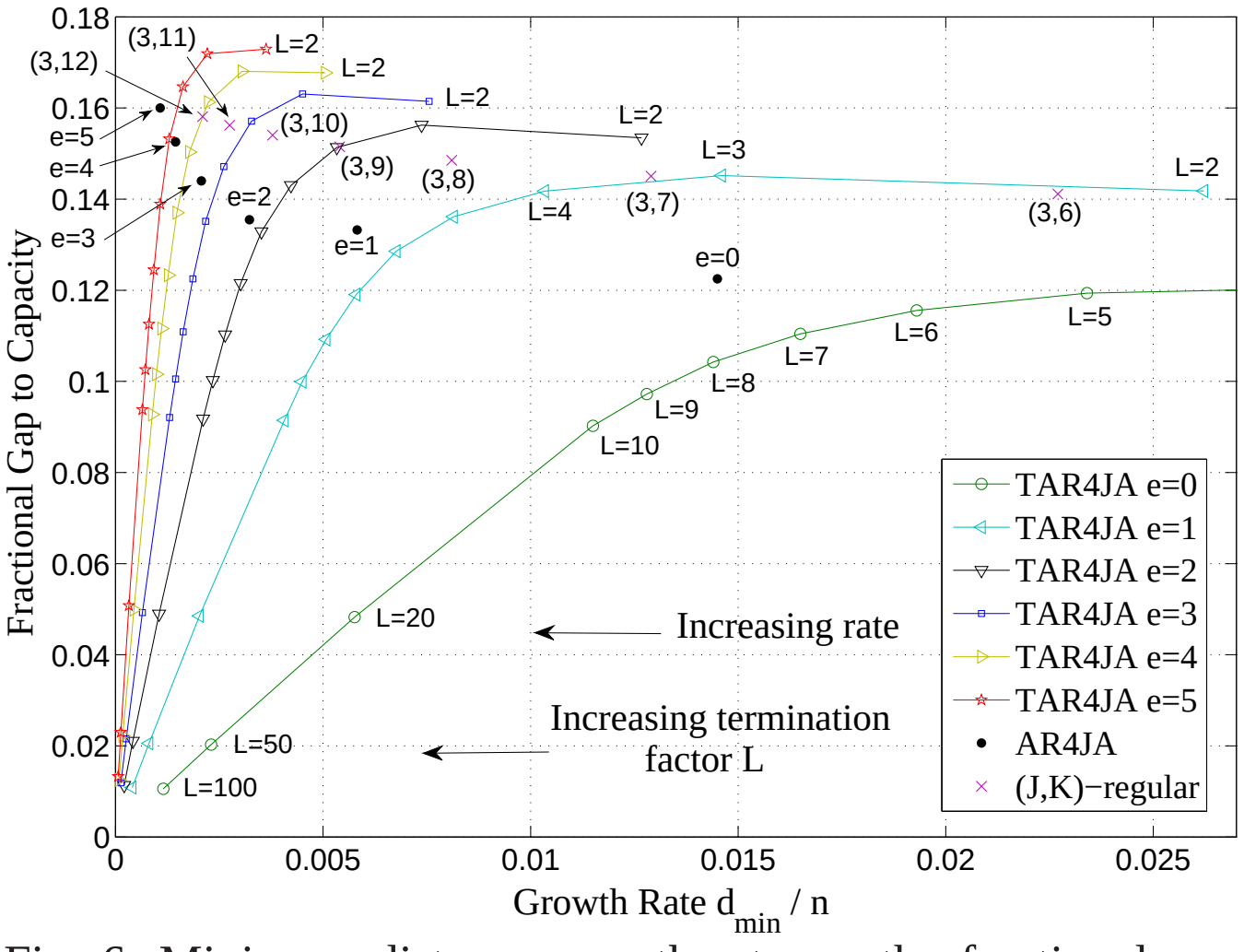

Fig. 6: Minimum distance growth rate vs. the fractional gap to capacity.

The trade-off we observe effectively allows a code designer to 'tune' between distance growth rate and threshold by choosing the parameters $e$ and $L$. We observe that, in particular, intermediate values of $L$ provide thresholds with a small gap to capacity while maintaining a reasonable distance growth rate with only a small loss in code rate. The complexity of the TAR4JA ensembles (measured by average node degrees) increases slowly with $L$ and approaches that of the AR4JA ensemble for a given extension parameter $e$. Further, as $L$ becomes sufficiently large for the scaled growth rates to converge, we observe that the gaps to capacity are approximately proportional to $L$ for all of the TAR4JA ensembles. For example, we obtain about a $10\%$ gap to capacity by terminating after $L = 9$ time instants; a $5\%$ gap after $L = 20$ time instants; a $2\%$ gap after $L = 50$ time instants; and a $1\%$ gap after $L = 100$ time instants. Finally, the extension parameter $e$ can be chosen, where a larger $e$ gives a higher code rate but a lower distance growth rate and greater complexity.

## V. Conclusions

We have provided a construction technique for a large family of asymptotically good AR4JA-based terminated LDPC convolutional code ensembles with thresholds close to capacity. The design method, based upon an extension parameter $e$ and a termination factor $L$, is very flexible, allowing the selection of asymptotically good LDPC code ensembles for a wide variety of rates. It was also shown that the ensemble average minimum distance growth rates are higher than those of other code ensembles with similar complexity. Further, as a result of the variable node degree design, we ensure fast convergence rates and thresholds close to capacity.

The discussion in this paper was limited to the BEC; however, as we have recently shown in [15], similar behaviour is observed for the additive white Gaussian noise channel. Finally, although the AR4JA protograph was used as an example, the same design method can be applied to other irregular protographs to construct asymptotically good LDPC code ensembles with varying code rates and thresholds close to capacity. In practice, the design parameter $L$ adds an additional degree of freedom to given block code designs. Starting from any LDPC block code, it is possible to derive periodically time-varying terminated convolutional codes that share the same encoding and decoding architecture for arbitrary $L$.

[2] Complexity is measured by average node degree. When comparing the TAR4JA ensembles to the AR4JA ensembles with equal extension parameters, the average variable node degree is equal for all $L$, but the average check node degree is less for the TAR4JA ensembles for finite $L$ because of the termination.